\documentclass[10pt]{article}

\begin{document}

\title{Brane-world models emerging from collisions of plane waves in $5D$}
\author{J. Ponce de Leon\thanks{E-mail: jponce@upracd.upr.clu.edu}\\ Laboratory 
of Theoretical Physics, Department of Physics\\ 
University of Puerto Rico, P.O. Box 23343, San Juan, \\ PR 00931, USA} 
\date{August 2003}

\maketitle
\begin{abstract}

We consider brane-world models embedded in a five-dimensional bulk spacetime 
with a large extra dimension and a cosmological constant. The cosmology in $5D$  
possesses ``wave-like" character in the sense that the metric coefficients in the 
bulk are assumed to have the form of plane waves propagating in the fifth dimension. 
We model the brane as the ``plane" of collision of waves propagating 
in opposite directions along the extra dimension. This plane is a jump 
discontinuity which presents the usual ${\bf Z}_2$ symmetry of brane models. 
The model reproduces the {\em generalized} Friedmann equation for the evolution on 
the brane, regardless of the specific details  in $5D$. Model solutions with 
spacelike extra coordinate show the usual {\em big-bang} behavior, while those 
with timelike extra dimension present a  {\em big bounce}.  This bounce is an  
genuine effect of a timelike extra dimension.  We argue that, based on our 
current knowledge,  models having a  large timelike extra dimension cannot  
be dismissed as mathematical curiosities in non-physical solutions. 
The size of the extra dimension is small today, but it is {\em increasing} if 
the universe is expanding with  acceleration. Also, the expansion rate of the 
fifth dimension can be expressed in a simple way through the 
four-dimensional ``deceleration" and Hubble parameters as $- q H$. 
These predictions could have important observational implications, 
notably   for the time variation of rest mass,  electric charge and 
the gravitational ``constant".  They hold for the three $(k = 0, + 1, - 1)$ 
models with arbitrary cosmological constant, and are independent of the signature 
of the extra dimension.

 \end{abstract}



\newpage

\section{Introduction}
Recently, there has been an increased interest in models where 
our four-dimensional universe is embedded in a higher-dimensional 
bulk spacetime having large extra dimensions. In the brane-world scenario, 
standard-model  fields are confined to a singular $3$-brane, which is 
identified with our observed four-dimensional spacetime, while gravity 
can propagate in the extra dimension as well.   

As a consequence of the confinement of matter fields to a $3$-brane, 
solutions in brane-world theory can be obtained by solving  the 
five-dimensional Einstein equations in the bulk and then  applying Israel's
 boundary conditions across the brane, which is assumed to have ${\bf{Z}}_{2}$ 
symmetry.

In cosmological solutions, one approach is to solve  
the equations in a Gaussian normal coordinate system based on our 
brane-universe, which is taken to be {\em fixed} at some constant value of 
the extra coordinate, say $y = const$ \cite{Binetruy1}-\cite{Vollick}. In an 
alternative  approach the brane is described as a domain wall {\em moving} in 
a five-dimensional bulk, which is the  $5D$ analog of the static 
Schwarzschild-anti-de 
Sitter spacetime \cite{Ida}-\cite{Dadhich} . 

These both approaches lead to similar results. In particular, they produce 
the  same evolution equation, a generalized FRW equation, for the scale factor. 
They represent the {\em same} spacetime but in different coordinates \cite{mukoyama2}. 
Indeed, there exists a coordinate transformation that brings the $5D$ 
line element of the static $Sch-AdS$ bulk used in \cite{Ida}-\cite{Dadhich} 
into the  bulk in Gaussian normal coordinates, with static fifth dimension, 
used in \cite{Binetruy1}-\cite{Vollick}. 

The condition  of {\em static} fifth dimension is useful because it allows the  
complete integration of the brane-world equations in $5D$, with no more 
assumptions than an equation of state. However, it is an unrealistic external 
restriction, not a requirement of the field equations. There is no physical 
reason why the ``size" of the extra dimension  should not change during 
the evolution of the universe. It is therefore interesting to analyze 
brane-world models without this condition. Possible implications of a 
non-static extra dimension include the variation of fundamental 
physical ``constants" \cite{NewVar}. 

Here we present an exact model where the singular nature of the  brane, assumed 
fixed at $y = const$, comes out  in a natural way. Namely, we assume that the 
metric functions in $5D$ are plane waves moving along the extra dimension. We refer 
here to the case where the metric functions have a simple functional dependence 
of time and the extra coordinate similar to  that in traveling waves or pulses 
propagating in the fifth dimension. In a ${\bf Z}_{2}$-symmetric universe, 
these waves  propagate, with equal but opposite speed along $y$,  and 
collide at $y = 0$. The ``plane" of collision is a jump discontinuity 
that we identify  with our brane, and the material emerging from the 
(collision) discontinuity is described as an ``effective" matter in $4D$. 

The assumption of plane waves or ``wave-like" solutions in five-dimensions, 
similar to the one considered here, has previously been used  in the 
literature, although in another context, by Wesson, Liu and 
Seahra \cite{LiuWesson2}-\cite{LiuWesson3}, Horowitz, Low and 
Zee \cite{HLZ} as well as  the present author \cite{FRLW}. 

We will see that our model reproduces the generalized 
Friedmann  equation for the 
evolution on the brane, regardless of the specific details of 
the cosmology in $5D$. We will show that models with timelike 
extra dimension present a {\em big bounce}, in contrast with 
a {\em big bang} in models with spacelike extra dimension, 
where the geometry is regular and the matter quantities are finite. 
This bounce is not a consequence of a repulsive cosmological term, 
but is an  authentic effect of a timelike extra dimension. 

In the case of static extra dimension, the geometry  on the 
brane corresponds to Milne's universe. We use it here to 
illustrate two important points. Firstly,  from a technical 
viewpoint it illustrates the fact that a solution which looks 
complicated in $5D$, depending on many parameters, might 
correspond to a very simple solution on the brane, regardless 
of the specific choice of parameters in $5D$. Secondly, from a 
physical viewpoint it provides another example where  the  
brane-world paradigm leads to variable cosmological term and 
gravitational coupling to brane matter.

In addition, our plane-wave model allows us to make  
specific predictions regarding the extra dimension. We 
will show that (i) although the extra dimension is small 
today, it is increasing if the universe is expanding with  
acceleration, (ii) the expansion rate of the fifth dimension 
can be expressed in a simple way through the four-dimensional 
``deceleration" and Hubble parameters as $- q H$. These 
predictions  hold for the three $(k = 0, + 1, - 1)$ models and 
arbitrary cosmological constant. They could have important observational 
implications, notably   for the time variation of rest mass,  electric 
charge and the gravitational ``constant".

This paper is organized as follows. In the next Section we present 
the general $5D$-equations in the bulk as well as their solutions for 
the plane-wave model. In Section 3 we will see that the (generalized) 
Friedmann equation is recovered on the brane and analyze the evolution of 
models with timelike and spacelike extra dimension. In Section 4, we 
examine the behavior of the extra dimension. In Section 5, we critically 
review some objections commonly raised against the timelike signature of 
the extra coordinate. Finally, in Section 6 we give a summary.  

\section{Equations in the bulk}

For cosmological applications, we take metric in $5D$ as 
\begin{equation}
\label{cosmological metric}
d{\cal{S}}^2 = n^2(t,y)dt^2 - a^2(t,y)\left[\frac{dr^2}{(1 - kr^2)} 
+ r^2(d\theta^2 + \sin^2\theta d\phi^2)\right] + \epsilon \Phi^2(t, y)dy^2,
\end{equation}
where $k = 0, +1, -1$ and $t, r, \theta$ and $\phi$ are the 
usual coordinates for a spacetime with spherically symmetric 
spatial sections. We adopt signature $(+ - - - )$ for spacetime 
and the factor $\epsilon $ can be $- 1$ or $+ 1$ depending on whether 
the extra dimension is spacelike or timelike, respectively.
The corresponding field equations in $5D$ are 
\begin{equation}
\label{equations in 5D}
G_{AB} = k_{(5)}^2 {^{(5)}T}_{AB},
\end{equation}
where $k_{(5)}^2$ is a constant introduced for dimensional 
considerations, ${^{(5)}T}_{AB}$ is the energy-momentum tensor in $5D$ and  
the non-vanishing components of the Einstein tensor $G_{AB}$ are
\begin{equation}
\label{G 00}
G_{0}^{0} = \frac{3}{n^2}\left(\frac{{\dot{a}}^2}{a^2} + 
\frac{\dot{a}\dot{\Phi}}{a \Phi}\right) + \frac{ 3 \epsilon}{\Phi^2}\left(\frac{a''}{a}
 + \frac{{a'}^2}{a^2} - \frac{a' \Phi'}{a \Phi}\right) + \frac{3 k}{a^2},
\end{equation}
\begin{eqnarray}
\label{G 11}
G^{1}_{1} = G^{2}_{2} = G^{3}_{3} &=& \frac{1}{n^2}\left[\frac{\ddot{\Phi}}{\Phi}
 + \frac{2\ddot{a}}{a} + \frac{\dot{\Phi}}{\Phi}\left(\frac{2 \dot{a}}{a} 
- \frac{\dot{n}}{n}\right) + \frac{\dot{a}}{a}\left(\frac{\dot{a}}{a} 
- \frac{2 \dot{n}}{n}\right)\right] + \nonumber \\
& & \frac{\epsilon}{\Phi^2}\left[\frac{2 a''}{a} + \frac{n''}{n} 
+ \frac{a'}{a}\left(\frac{a'}{a} + \frac{2 n'}{n}\right) 
- \frac{\Phi'}{\Phi}\left(\frac{2a'}{a} + \frac{n'}{n}\right)\right] + \frac{k}{a^2},
\end{eqnarray}
\begin{equation}
\label{G zero four}
G^{0}_{4} = \frac{3}{n^2}\left(\frac{{\dot{a}}'}{a} - \frac{\dot{a} n'}{a n}
 - \frac{a' \dot{\Phi}}{a \Phi}\right),
\end{equation}
and 
\begin{equation}
\label{G 44}
G_{4}^{4} = \frac{3}{n^2}\left(\frac{\ddot{a}}{a} + \frac{{\dot{a}}^2}{a^2}
 - \frac{\dot{a}\dot{n}}{a n}\right) + \frac{ 3 \epsilon}{\Phi^2}\left(\frac{{a'}^2}{a^2} 
+ \frac{a' n'}{a n}\right) + \frac{3 k}{a^2}. 
\end{equation}
Here a dot and a prime denote partial derivatives with 
respect to $t$ and $y$, respectively. 

Introducing the function \cite{Binetruy2}
\begin{equation}
\label{F}
F(t,y) = k a^2 + \frac{(\dot{a}a)^2}{n^2} + \epsilon \frac{(a' a)^2}{\Phi^2},
\end{equation}
which is a first integral of the field equations, we get  
\begin{equation}
\label{F prime}
F' = \frac{2a' a^3}{3}k_{(5)}^2 {^{(5)}T}^{0}_{0},
\end{equation}
and
\begin{equation}
\label{F dot}
\dot{F} = \frac{2\dot{a} a^3}{3}k_{(5)}^2 {^{(5)}T}^{4}_{4}.
\end{equation}

In what follows we will assume that the five-dimensional energy-momentum 
tensor has the form
\begin{equation}
\label{AdS}
{^{(5)}T}_{AB} =  \Lambda_{(5)}g_{AB}, 
\end{equation}
where $\Lambda_{(5)}$ is the cosmological constant in the bulk. It 
can be (i) positive as in the usual de Sitter $(dS_{5})$ solution, 
(ii) negative as in the brane-world scenarios where our spacetime 
is identified with a singular hypersurface (or $3$-brane) embedded 
in an $AdS_{5}$ bulk, or (iii) zero as in STM where the matter in $4D$ is 
interpreted as an effect of the geometry in $5D$.

\subsection{Plane waves along the fifth dimension}

We now  assume that the metric coefficients in (\ref{cosmological metric}) 
are ``wave-like" functions of the argument $(t - \lambda y)$:
\begin{equation}
n = n(t - \lambda y), \;\;\;a = a(t - \lambda y), \;\;\; \Phi = \Phi(t - \lambda y),
\end{equation}
where $\lambda$ can be interpreted  as the ``wave number".  
Now, from $G^{0}_{4} = 0$ we get
\begin{equation}
\label{a dot in terms of n and Phi}
\dot{a} = \alpha n \Phi,
\end{equation}
where $\alpha$ is a constant of integration. Substituting this into (\ref{F}) we obtain
\begin{equation}
\label{definition of f}
F = a^2\left(k + \alpha^2 \Phi^2 + \epsilon \lambda^2 \alpha^2 n^2\right) \equiv a^2 f^2.
\end{equation}
The auxiliary function $f$ satisfies the equation 
\begin{equation}
a f \frac{df}{da} + f^2 = \frac{a^2}{3}k_{(5)}^2 \Lambda_{(5)},
\end{equation}
which follows from (\ref{F dot}), (\ref{AdS}) and (\ref{definition of f}). 
Integrating we get 
\begin{equation}
f^2 = \frac{\beta \alpha^2}{a^2} + \frac{1}{6}a^2 k_{(5)}^2 \Lambda_{(5)}.
\end{equation}
Consequently, 
\begin{equation}
\label{Phi}
\Phi^2 = - \epsilon \lambda^2 n^2 - \frac{k}{\alpha^2} + \frac{\beta}{a^2} + \frac{a^2}{6 \alpha^2} k_{(5)}^2 \Lambda_{(5)},
\end{equation}
and  from (\ref{a dot in terms of n and Phi})
\begin{equation}
\label{relation between a and n}
\left(\frac{\dot{a}}{n}\right)^2 + k = - \epsilon \lambda^2 \alpha^2 n^2 
+ \frac{\beta \alpha^2}{a^2} + \frac{a^2}{6}k_{(5)}^2 \Lambda_{(5)}.
\end{equation}
After some manipulations one can verify that the remaining field 
equation $G_{1}^{1} = G_{2}^{2} = G_{3}^{3} = k_{(5)}^2 \Lambda_{(5)}$ 
is identically satisfied.

Thus, the complete specification of the solution requires the consideration 
of some physics, or a simplifying mathematical assumption, to 
determine $ \dot{a}$ (or $n$). Then, from (\ref{relation between a and n}) 
we find $n$ (or $a$). Finally, the function $\Phi$ is given by  (\ref{Phi}). The 
whole solution, thus specified, depends on three 
parameters, $\alpha$, $\beta$ and $\lambda$. 

\subsection{Generality of the wave-like solutions}

Simple power-law solutions to the above equations have been discussed 
by Wesson, Liu and Seahra in another 
context \cite{LiuWesson2}, \cite{LiuWesson3}. They considered a 
spacelike extra dimension and made some simplifying assumptions, 
among others that $\Lambda_{(5)} = k = \beta = 0$.

The natural question to ask is whether the plane-wave model 
in not too restrictive as to prevent the existence of more general 
wave-like solutions, under less restrictive assumptions. 

Before going on with our study, we should elucidate  this question. We 
demonstrate here that the answer to this question is negative. Namely, we show 
a simple class of wave-like solutions, which impose no restrictions on the 
parameters $\Lambda_{(5)}$, $k$, $\beta$ or the signature of the extra dimension.

In order to complete the set of equations (\ref{Phi}), (\ref{relation between a and n}) 
we consider $\dot{\Phi} = 0$. Then,  without loss of generality we can set 
\begin{equation}
\label{Static fifth dimension}
\Phi = 1.
\end{equation}
The solution to the above equations for a spacelike extra dimension $(\epsilon = -1)$ is
\begin{equation}
\label{spacelike extradimension}
a^2(t,y) = A \cosh\,[ \frac{\omega}{\lambda}(t - \lambda y)] + B \sinh\, 
 [\frac{\omega}{\lambda}(t - \lambda y)] - \frac{2[k + \alpha^2]}{\omega^2}.
\end{equation}
While for a timelike extra dimension $(\epsilon = 1)$,
\begin{equation}
\label{timelike extradimension}
a^2(t,y) = A \cos\, [ \frac{\omega}{\lambda}(t - \lambda y)] +
 B \sin\, [ \frac{\omega}{\lambda}(t - \lambda y)] - \frac{2[k + \alpha^2]}{\omega^2}.
\end{equation}
where
 \begin{equation}
\omega = \sqrt{\left( - \frac{2}{3}k_{(5)}^2 \Lambda_{(5)}\right)},
\end{equation}
and the constants  $A$ and  $B$ are related by
\begin{equation}
\epsilon B^2 =  - A^2 + \frac{4 \alpha^2}{\omega^2}\left[ \beta + 
\frac{(k + \alpha^2)^2}{\alpha^2 \omega^2}\right].
\end{equation} 
We note that for a  de Sitter bulk $(dS_{5})$ the situation is reversed. Namely, 
for a spacelike extra dimension the time-evolution of the brane in a $dS_{5}$ bulk is 
given by (\ref{timelike extradimension}), while for a timelike  by (\ref{spacelike 
extradimension}). 

The above solution proves that the plane-wave model is not ``too" restrictive, but  
it is compatible with a wide range of parameters. Other general solutions of this 
kind exist, but we will not discuss them here. What we will discuss is the plane-wave 
model in the context of the brane-world paradigm.    

\section{The brane in a ${\bf{Z}}_{2}$-symmetric bulk}

The scenario in brane-world models is that matter fields are 
confined to a singular $3$-brane. We now  proceed to construct such a brane. 
For convenience the coordinate $y$ is chosen such that the 
hypersurface $\Sigma: y = 0$ coincides with the brane, which 
is assumed to be ${\bf Z}_2$ symmetric in the 
bulk background  \cite{Randall1}-\cite{Maartens2}. The brane 
is obtained by a simple ``copy and paste" procedure. Namely, 
we cut the generating $5D$ spacetime, in two pieces along $\Sigma$, 
then copy the region $y \leq 0$ and paste it in the region $y \geq 0$. 
The result is a singular hypersurface in a ${\bf Z}_2$ symmetric universe 
with metric
\begin{equation}
\label{Bulk +}
d{\cal{S}}^2 = n^2(t + \lambda y)dt^2 - a^2(t + \lambda y)\left[\frac{dr^2}{(1 - kr^2)} 
+ r^2(d\theta^2 + \sin^2\theta d\phi^2)\right] + \epsilon \Phi^2(t + \lambda y)dy^2,
\end{equation}
\begin{equation}
\label{Bulk -}
d{\cal{S}}^2 = n^2(t - \lambda y)dt^2 - a^2(t - \lambda y)\left[\frac{dr^2}{(1 - kr^2)} 
+ r^2(d\theta^2 + \sin^2\theta d\phi^2)\right] + \epsilon \Phi^2(t - \lambda y)dy^2,
\end{equation}
for $y > 0 $ and $y  < 0$, respectively. They can be interpreted as plane-waves 
propagating in  ``opposite" directions along the fifth dimension, 
and colliding at $y = 0$. 

If we introduce the normal unit ($n_{A}n^{A} = \epsilon$) vector, orthogonal to hypersurfaces $y = constant$
\begin{equation}
n^A = \frac{\delta^{A}_{4}}{\Phi}\;  , \;\;\;\;\;  n_{A}= (0, 0, 0, 0, \epsilon \Phi),
\end{equation}
then the extrinsic curvature $K_{\mu\nu}$ is  
\begin{equation}
\label{extrinsic curvature}
K_{\alpha\beta} = \frac{1}{2}{\cal{L}}_{n}g_{\alpha\beta} = 
\frac{1}{2\Phi}\frac{\partial{g_{\alpha\beta}}}{\partial y},\;\;\; K_{A4} = 0.
\end{equation} 
The metric is continuous at $y = 0$, but $K_{\mu\nu}$ is not. The jump of $K_{\mu\nu}$ is related to $\tau_{\mu\nu}$, the energy-momentum tensor of the matter on the brane, through Israel's boundary conditions, viz.,
\begin{equation}
\label{boundary conditions}
K_{\mu\nu}\mid_{{\Sigma}^{+}} - K_{\mu\nu}\mid_{{\Sigma}^{-}} = 
- \epsilon k_{(5)}^2 \left({\tau}_{\mu\nu} - \frac{1}{3}g_{\mu\nu}\tau\right).
\end{equation}
Thus, 
\begin{equation}
\label{K in terms of S}
K_{\mu\nu}\mid_{{\Sigma}^{+}} =  - K_{\mu\nu}\mid_{{\Sigma}^{-}}
 = - \frac{\epsilon}{2}k_{(5)}^2 \left({\tau}_{\mu\nu} - \frac{1}{3}g_{\mu\nu}\tau\right).
\end{equation}
Consequently, 
\begin{equation}
\label{emt on the brane in terms of K}
\tau_{\mu\nu} = - \frac{2\epsilon}{k_{(5)}^2}\left(K_{\mu\nu} - g_{\mu\nu} K\right).
\end{equation}
From the field equation $G_{04} = k_{(5)}^2 {^{(5)}T}_{04}$ and (\ref{AdS}) 
it follows that
\begin{equation}
\label{conservation of emt on the brane}
\tau^{\mu}_{\nu;\mu} = 0.
\end{equation}
Thus $\tau_{\mu\nu}$ represents the total conserved energy-momentum 
tensor on the brane. It is usually separated in  two parts, 
\begin{equation}
\label{decomposition of tau}
\tau_{\mu\nu} =  \sigma g_{\mu\nu} + T_{\mu\nu},
\end{equation} 
where $\sigma$ is the tension of the brane in  $5D$, and $T_{\mu\nu}$ 
represents the energy-momentum tensor of ordinary matter in $4D$. Finally, 
from (\ref{K in terms of S}), (\ref{emt on the brane in terms of K}) 
and (\ref{decomposition of tau}) we get
\begin{equation}
\label{K in terms of matter in the brane}
K_{\mu\nu}\mid_{{\Sigma}^{+}} = -  \frac{\epsilon k_{(5)}^2}{2} \left(T_{\mu\nu} 
- \frac{1}{3}g{\mu\nu}(T + \sigma)\right).
\end{equation}

\subsection{Equation of state}

Here, in order to complete the system of equations (\ref{Phi}), 
(\ref{relation between a and n}) we make some assumptions on the character
 of the matter on the brane. In cosmological applications, the ordinary 
matter is usually assumed to be  a perfect fluid 
\begin{equation}
T_{\mu\nu} = (\rho + p)u_{\mu}u_{\nu} - p g_{\mu\nu},
\end{equation}
where the  energy density $\rho$ and pressure $p$ satisfy the 
isothermal equation of state, viz.,
\begin{equation}
\label{equation of state}
p = \gamma \rho, \;\;\;\  0 \leq \gamma \leq 1.
\end{equation}
From (\ref{extrinsic curvature}) and (\ref{Bulk +}) we find
\begin{eqnarray}
\label{K +}
K^{t}_{t} &=& \frac{\lambda \dot{n}}{n \Phi},\nonumber \\
K^{r}_{r} &=& K^{\theta}_{\theta} = K^{\phi}_{\phi} = \frac{\lambda \dot{a}}{a \Phi}.
\end{eqnarray}
Using (\ref{Bulk -}) we obtain the same expressions as in (\ref{K +}) 
but with opposite sign, as one expected. Thus, 
using (\ref{K in terms of matter in the brane}) 
and (\ref{equation of state}) we obtain 
\begin{equation}
\label{density}
\rho(t) = \frac{2 \epsilon \lambda}{k_{(5)}^2 (\gamma + 1)\Phi|_{brane}}\left[\frac{\dot{a}}{a} - \frac{\dot{n}}{n}\right]_{brane},
\end{equation}
and 
\begin{equation}
\label{tension of the brane}
\sigma =   
\frac{2 \epsilon \lambda}{k_{(5)}^2 (\gamma + 1)\Phi|_{brane}}
\left[(3\gamma + 2)\frac{\dot{a}}{a} + \frac{\dot{n}}{n} \right]_{brane}.
\end{equation}
Using (\ref{a dot in terms of n and Phi}) the last equation can be written as  
\begin{equation}
\label{relation between n, a and sigma}
\frac{dn}{da} + (3\gamma + 2)\frac{n}{a} = 
\frac{\epsilon \sigma k_{(5)}^2(\gamma + 1)}{2 \lambda \alpha}, 
\end{equation}
which can be easily integrated if the tension $\sigma$ is assumed to be constant, viz., 
\begin{equation}
\label{n through a}
n = \frac{C}{a^{3\gamma + 2}}+ \frac{\epsilon \sigma k_{(5)}^2}{6 \lambda \alpha}a,
\end{equation}
where $C$ is an integration constant. The solution in the bulk is now 
fully specified; substituting this expression into (\ref{Phi}) we get $\Phi$ 
as a function of $a$. Then from (\ref{relation between a and n}) 
and (\ref{n through a}), we obtain the differential equation for 
the scale factor $a$. This  completes the solution.  

\subsection{The solution on the brane}

The specific form of the solution in the bulk is rather cumbersome. Indeed, 
the function $\Phi = \Phi(a)$ and the differential equation 
for $a = a(t - \lambda y)$ are difficult to manage due to the number 
of terms and parameters in the equations. 

However, if we are interested in $4D$ we do not have to worry 
about the many details of the solutions in $5D$. For all we need are 
the equations expressed  in terms of the cosmological (or proper) 
time $\tau_{\Sigma}$ on the brane and not in terms of the coordinate 
time  in $5D$. Thus, using  
\begin{equation}
\label{proper time}
d{\tau}_{\Sigma} = n(t, 0)dt,
\end{equation}
the metric on the brane fixed at $y = 0$ becomes
\begin{equation}
ds^2 = (d{\tau}_{\Sigma})^2 - a^2(\tau_{\Sigma})\left[\frac{dr^2}{(1 - kr^2)}
 + r^2(d\theta^2 + \sin^2\theta d\phi^2)\right], 
\end{equation}
where the scale factor is given by\footnote{The most general brane-Universe 
solutions for a three-brane in a five dimensional spacetime have 
been found in \cite{BCG}.}
\begin{equation}
\label{scale factor 1}
\left(\frac{da}{d\tau_{\Sigma}}\right)^2 = -  k + 
\frac{a^2 k_{(5)}^2}{6}\left(\Lambda_{(5)} - \epsilon \frac{k_{(5)}^2 \sigma^2}{6}\right) 
+ \frac{\beta \alpha^2}{a^2} - \frac{C \lambda \alpha \sigma k_{(5)}^2}{3 a^{3 \gamma +1}}
 - \frac{\epsilon C^2 \lambda^2 \alpha^2}{a^{6\gamma + 4}}.
\end{equation}
For the energy density we find
\begin{equation}
\rho = \frac{6 \epsilon C \alpha \lambda}{k_{(5)}^2 a^{3(\gamma + 1)}}.
\end{equation} 
Clearly we should require $(\epsilon C \alpha \lambda) > 0$. If we make the 
identification 
\begin{equation}
\label{definition of lambda}
\Lambda_{(4)} = \frac{1}{2}k_{(5)}^2\left(\Lambda_{(5)} - 
\epsilon \frac{ k_{(5)}^2 \sigma^2}{6}\right),
\end{equation}
\begin{equation}
\label{effective gravitational coupling}
8 \pi G =  - \epsilon \frac{k_{(5)}^4 \sigma}{6},
\end{equation}
then (\ref{scale factor 1}) becomes
\begin{equation}
\label{generalized FLRW equation}
\frac{3}{a^2}\left(\frac{da}{d\tau_{\Sigma}}\right)^2 + \frac{3k}{a^2} 
= \Lambda_{(4)} + 8\pi G \rho - \epsilon \frac{\rho^2}{12} 
+ \frac{3 \beta \alpha^2}{a^4}.
\end{equation}

\medskip

We note that the parameter $\beta$ is related to the so-called Weyl or black radiation. 
Namely,  the projection of the bulk Weyl tensor ${^{(5)}C}_{ABCD}$ 
orthogonal to ${\hat{n}}^A$, i.e., ``parallel" to spacetime, is given by
\begin{eqnarray}
\label{Weyl Tensor}
E_{\alpha\beta} &=& {^{(5)}C}_{\alpha A \beta B}n^An^B\nonumber \\
&=& - \frac{1}{\Phi}\frac{\partial K_{\alpha\beta}}{\partial y} 
+ K_{\alpha\rho}K^{\rho}_{\beta} - \epsilon \frac{\Phi_{\alpha;\beta}}{\Phi}
 - \epsilon \frac{k^{2}_{(5)}}{3}\left[{^{(5)}T}_{\alpha\beta} + ({^{(5)}T}^{4}_{4}
 - \frac{1}{2}{^{(5)}T})g_{\alpha\beta}\right].
\end{eqnarray}
Substituting the above solution into this expression we obtain 
\begin{equation}
\label{black radiation}
8 \pi G \rho_{Weyl} = - \epsilon E_{0}^{0} 
= \frac{3 \beta \alpha^2}{a^4}, \;\;\; p_{Weyl} = \frac{1}{3}\rho_{Weyl}, 
\end{equation}
where $8 \pi G p_{Weyl} = \epsilon E^{1}_{1} = \epsilon E^{2}_{2} = \epsilon E^{3}_{3}$. 
Thus setting $\beta = 0$ is equivalent to eliminating the 
contribution coming from the free gravitational field. 

With  (\ref{definition of lambda}) and  (\ref{effective gravitational coupling}) 
as the definitions of the fundamental quantities $\Lambda_{(4)}$ and $G$, 
the evolution equation (\ref{generalized FLRW equation}) is 
the ``generalized" Friedmann equation. It reduces to the usual one of  general relativity
for $\sigma >> \rho$ and contains  higher-dimensional 
modifications to general relativity. Namely, local quadratic energy-momentum 
corrections via  $\rho^2$, and the nonlocal effects from the free 
gravitational field in the bulk, transmitted  by $E_{\mu\nu}$. 

\subsubsection{Spatially flat universe with $\Lambda_{(4)} = 0$}

This is an important case because astrophysical data, including the 
age of the universe, are compatible with cosmological models with flat $(k = 0)$, 
space sections. On the other hand the bulk cosmological constant is commonly 
chosen so that the brane cosmological constant vanishes. 

For simplicity, let us first consider $\beta = 0$. We recall that the term 
associated with $\beta$ is related to the bulk Weyl tensor,  which is 
constrained to be small enough at the time of nucleosynthesis and it should 
be negligible today.  

For this case the evolution equation (\ref{scale factor 1}) can be exactly 
integrated for any value of $\gamma$ in the equation of state 
(\ref{equation of state}), viz., 
\begin{equation}
\label{scale factor for beta, Lambda and k are zero} 
a^{3(\gamma + 1)}  = 
\frac{3 (\epsilon C \lambda \alpha)}{4 (- \epsilon \sigma)k_{(5)}^2}
\left[\sigma^2 k_{(5)}^4 (\gamma + 1)^2(\tau_{\Sigma} - {\bar{\tau}}_{\Sigma})^2 
+ 4 \epsilon\right],
\end{equation}
where ${\bar{\tau}}_{\Sigma}$ is a constant of integration. Now the 
energy density $\rho$ becomes
\begin{equation}
\label{density for beta, lambda and k zero}
\rho = \frac{8 (- \epsilon \sigma)}{\left[\sigma^2 k_{(5)}^4 (\gamma + 1)^2(\tau_{\Sigma}
 - {\bar{\tau}}_{\Sigma})^2 + 4 \epsilon \right]}.
\end{equation} 
In this model, the only restrictions on the constants $\alpha$, $C$, 
$\sigma$ and $\epsilon$ come from the positiveness of $G$ and $\rho$. 
Namely $- \epsilon \sigma >0$ and $\epsilon C \alpha \lambda >0$, which 
is equivalent to $\sigma C \alpha <0$ (we assume $\lambda >0$). These 
conditions assure that $a$ is positive at all times. 

It is interesting to note the role of the signature of the extra dimension. 

\medskip

\paragraph{Big bang:}For a spacelike extra dimension $(\epsilon = -1)$ we 
choose the constant ${\bar{\tau}}_{\Sigma}$ so that $a = 0$ at $\tau_{\Sigma} = 0$, 
for which $\rho \rightarrow \infty$. Then $G > 0$, requires positive brane 
tension $\sigma$ and $\Lambda_{(4)} = 0$ requires $\Lambda_{(5)} < 0$, that 
is a $AdS$-bulk.

\medskip

\paragraph{Big bounce:} The situation is different for a timelike 
extra dimension $(\epsilon = + 1)$. Here the scale factor never reaches 
zero. Instead there is a finite minimum 
for  $a$ at $\tau_{\Sigma} = {\bar{\tau}}_{\Sigma}$, before 
which the universe contracts and after which it expands. This minimum is 
\begin{equation}
\label{finite minimum}
\left(a_{min}\right)^{3(\gamma + 1)} = 
\frac{3(- \sigma C \alpha \lambda)}{\sigma^2 k_{(5)}^2} > 0.
\end{equation}
Consequently, the energy density does not diverge at $\tau_{\Sigma } 
= {\bar{\tau}}_{\Sigma}$. Its maximum value is
\begin{equation} 
\label{max rho}
\rho_{max} = (- 2 \sigma) >0.
\end{equation}
Here, positiveness of $G$ requires negative tension. On the other hand, 
$\Lambda_{(4)} = 0$, requires the bulk to 
be a $dS$-bulk, viz., $\Lambda_{(5)} > 0$.

\medskip

A similar exact integration can be performed for $\beta \neq 0$ in the 
radiation dominated era, $p = \rho/3$. The solution is obtained from the 
above by setting $\gamma = 1/3$ and substituting
\begin{equation}
 \sigma k_{(5)}^2 \rightarrow \sigma k_{(5)}^2 - \frac{3 \beta \alpha}{C \lambda}.
\end{equation}
We can see now the effects of $\beta$, on the minimum (\ref{finite minimum}) 
for $\epsilon = +1$. Namely
\begin{equation}
\left(\frac{a_{min}(\beta = 0)}{a_{min}(\beta)}\right)^4 = 
\left[ 1 + \frac{3 \beta \alpha^2}{(- \alpha C \lambda \sigma)}\right].
\end{equation}
Since $(- \alpha C \lambda \alpha) > 0$, the introduction of a positive 
$\beta$ pushes the minimum towards the ``origin", to the region 
$0 < a_{min}(\beta) < a_{min}(\beta = 0)$. While a negative $\beta$ 
moves it away from the origin,  to the region 
$a_{min}(\beta = 0) < a_{min}(\beta) < \infty$. At the intuitive 
level, this means that for a positive (negative)  $\beta$ the Weyl 
tensor tends to increase (decrease) the effective gravitational mass. 
This result is consistent with previous studies \cite{Old JPdeL}. 

\medskip

We note that a similar ``bouncing" behavior can happen in 
 general relativity in presence of a large, 
positive cosmological constant. The interesting fact is that here the 
cosmological constant $\Lambda_{(4)}$ is zero; in our model the bounce is 
produced by the timelike extra dimension 

\subsubsection{Milne vacuum universe}
We have already mentioned that the solution in the bulk can be very 
intricate, but very simple on the brane. A nice example of this is 
provided by the wave-like model  with  $\dot{\Phi} = 0$. 
The exact  solution for the scale factor in the bulk is given 
by (\ref{spacelike extradimension}), or (\ref{timelike extradimension}), 
and the expression for $n$ is obtained from (\ref{a dot in terms of n and Phi}). 
However, the solution on the brane  is straightforward. Indeed, from  
(\ref{a dot in terms of n and Phi}) and (\ref{proper time}) we get
\begin{equation}
a = \alpha (\tau_{\Sigma} - {\bar{\tau}}_{\Sigma}),
\end{equation}
This scale factor corresponds to the Milne vacuum universe, for 
which the total or effective energy density $\rho_{eff}$ and pressure $p_{eff}$ satisfy 
the equation of state $(\rho_{eff} + 3 p_{eff}) = 0$. This is also interpreted 
as the equation of state for non-gravitating matter\footnote{The equation 
of state $\rho = - 3 p$ appears in different contexts: in discussions 
of premature recollapse problem \cite{Barrow}, in coasting 
cosmologies \cite{Kolb}, in cosmic strings \cite{Vilenkin}, \cite{Gott}, in 
derivations of four-dimensional matter from the 
geometry in $5D$ \cite{Davidson}, \cite{space} and in 
limiting configurations \cite{Limit}}.

\paragraph{Variable $\sigma$:}An interesting feature here is that the tension 
of the brane is {\em not} a constant, but a function of time. Indeed, now $n$ 
satisfies (\ref{Phi}), with $\Phi = 1$, instead of (\ref{n through a}) which
 we integrated  for $\sigma = const$. Therefore, if  now 
substitute  (\ref{spacelike extradimension}), or (\ref{timelike extradimension}), 
and $n = \dot{a}/\alpha$ into (\ref{tension of the brane}) we obtain 
a function $\sigma = \sigma(a)$ and not $\sigma = const$. 

\paragraph{Variable $G$ and $\Lambda_{4}$:}Consequently, the 
fundamental quantities $G$ and $\Lambda_{(4)}$, for the case 
under consideration  are variable. This does not contradict other 
integrations \cite{Binetruy2}  with static $\Phi$ where the time 
dependence is determined from the boundary conditions for constant
 $\sigma$. What happens here is that the assumptions of plane-wave 
plus  $\Phi = 1$ simply leave no room for a constant $\sigma$. 
Different scenarios for the variation of $G$ and $\Lambda_{(4)}$, 
in the context of the brane-world paradigm, are discussed in Ref. \cite{NewVar}.

\medskip

We would like to finish this section with the following comments:

\medskip

(i) For $\lambda = 0$, the metric of the five-dimensional bulk is 
independent of the extra coordinate. Consequently, the brane becomes 
devoid of ordinary matter and there is only Weyl radiation and $\Lambda_{(4)}$. 

(ii) For $C = 0$ the brane is empty again, although the
 metric {\em does} depend on the extra dimension. Thus, a 
non-trivial dependence of the bulk metric on the extra coordinate 
is a necessary, {\em but  not a sufficient}, condition for the
 brane not to be empty.

(iii) In many papers the function $n(t,y)$ is subjected 
to the boundary condition  $n(t,0) = 1$. We notice that in our case it 
is not possible to impose this condition. What we do on the brane is to 
define the proper time as in (\ref{proper time}).  

\section{Evolution of the extra dimension}
We have seen that the solution on the brane makes no reference to the extra 
dimension, except for its signature. The explicit form of $\Phi$ enters 
nowhere in the discussion. However the model predicts a specific behavior 
for $\Phi$.  Indeed, from (\ref{a dot in terms of n and Phi}), 
(\ref{proper time}) and (\ref{generalized FLRW equation}) we have 
\begin{equation}
\label{turning points}
\alpha^2 \Phi^2 = - k + \frac{\Lambda_{(4)}a^2}{3} + \frac{8\pi G \rho a^2}{3} 
- \epsilon \frac{\rho^2 a^2}{36} + \frac{ \beta \alpha^2}{a^2} 
=  \left(\frac{da}{d\tau_{\Sigma}}\right)^2.
\end{equation}
Thus at  the ``turning" points $\Phi = 0$. These appear in closed $k = + 1$ 
and bouncing models. Since 
\begin{equation}
\alpha \left(\frac{d \Phi}{d \tau_{\Sigma}}\right) = \frac{d^2 a}{d\tau_{\Sigma}^2},
\end{equation}
it follows that $\alpha \Phi$ monotonically decreases (increases) if the 
universe is speeding down (up) during its 
expansion\footnote{The coefficient $\alpha$ can be absorbed 
into $\Phi$ by a simple change of scale in $y$.}. 
For the model discussed in Section 3.2.1.   
\begin{equation}
\label{Phi for the sol}
\alpha^2 \Phi^2 = \frac{k_{(5)}^2 (- \sigma C \alpha \lambda)}{3 a^{3\gamma + 1}} 
- \epsilon \frac{ C^2 \lambda^2 \alpha^2 }{k_{(5)}^4 a^{6\gamma + 4}},
\end{equation}
the extra dimension is small at the present dust dominated era, namely
\begin{equation}
\label{approx behavior of Phi}
\alpha \Phi \approx \frac{2}{3\tau_{\Sigma}},
\end{equation}
where we have used (\ref{scale factor for beta, Lambda and k are zero} ) 
and (\ref{density for beta, lambda and k zero}) with $\gamma = 0$. This is 
independent of the signature of the extra dimension. 

However, at early stages the behavior is different. Namely, for a 
spacelike extra coordinate, $\Phi$ decreases monotonically from a value 
that is formally infinite at the big bang $(a = 0)$. Instead, for a timelike
 extra dimension $\Phi$ grows from zero at the time of 
bounce $\tau_{\Sigma} = {\bar{\tau}}_{\Sigma}$ until it 
reaches a maximum value, after which  it decreases asymptotically 
to zero as shown in (\ref{approx behavior of Phi}). 

Finally we note that the relative variation of $\Phi$ has a 
universal character, namely,
\begin{equation}
\label{rate of change of Phi}
\frac{1}{\Phi}\left(\frac{d \Phi}{d\tau_{\Sigma}}\right) = -  q H,
\end{equation}
where $H$ and $q$ are the Hubble and deceleration $4D$ parameters, respectively
\begin{equation}
\label{definition of H and q}
H = \frac{1}{a}\left(\frac{da}{d\tau_{\Sigma}}\right), \;\;\;q 
= - a \left(\frac{d^2 a}{d\tau_{\Sigma}^2}\right)\left(\frac{da}{d \tau_{\Sigma}}\right)^{- 2}.
\end{equation}
According to modern observations, the universe is expanding with an acceleration, 
so that the parameter $q$ is, roughly, $- 0.5 \pm 0.2$. 
Assuming  $H = h \times 10^{-10} yr^{-1}$ we come to the 
estimate
\begin{equation}
\frac{1}{\Phi}\left(\frac{d\Phi}{d\tau_{\Sigma}}\right) 
\approx (3.5 \pm 1.4)\times 10^{-10} yr^{-1},
\end{equation}
where we have taken $h = 0.7$ \cite{Melnikov2}. This estimate holds for the 
three cases, $k = - 1, 0, + 1 $ and arbitrary cosmological constant, 
and set of parameters $(\beta, \sigma, \lambda, \alpha, C)$. It  
could have important observational implications because  it appears in 
different contexts, notably  in expressions concerning the variation of 
rest mass \cite{DynKK},  electric charge \cite{QMKK} and variation of 
the gravitational ``constant" $G$ \cite{Melnikov3},\cite{Melnikov1}.  

In general, the effective rest mass measured in $4D$ changes as 
the test particle  travels on $5D$ geodesics\footnote{The general, invariant
 equations for the change of mass are given in \cite{DynKK}.}. The total 
change consists of two parts, one of them is induced by the non-trivial 
dependence of the metric on the extra coordinate $(\partial g_{\mu\nu}/\partial y \neq 0)$
 and the other part is due to $\dot{\Phi}/\Phi$. Even in the simplest situation, 
where the metric does not depend on the extra coordinate, but only on 
time, $m_{0}$ the effective rest mass  in $4D$ of a  massless particle 
in $5D$ would change as
\begin{equation}
\frac{1}{m_{0}}\frac{dm_{0}}{d\tau_{\Sigma}} 
= - \frac{1}{\Phi}\frac{d\Phi}{d\tau_{\Sigma}},
\end{equation}
where we have used equation (25) in Ref. \cite{DynKK}. 
Similarly, the variation of $\Phi$ induces a  change in the  
electric charge, and consequently in the fine structure constant \cite{QMKK}.  

Regarding the time-variation of $G$, it is remarkable that 
in different  models with extra dimensions the ratio $(\dot{G}/G)$ 
 is found  to be proportional to $(\dot{\Phi}/\Phi)$ \cite{NewVar},\cite{Melnikov3}.
 At this point we have to mention that the specific value of $(\dot{\Phi}/\Phi)$ 
depends on the cosmological model. For example, for the cosmologies with 
separable metric coefficients \cite{JPdeL 1} $(\dot{\Phi}/\Phi) = (1 + q)H$, 
instead of (\ref{rate of change of Phi}).  Consequently, the measurement 
of quantities like $(\dot{m_{0}}/m_{0})$ and $(\dot{G}/G)$ will give 
the opportunity to test different models for compatibility with 
observational data.

\section{Signature of the extra dimension}

In both, compactified and non-compactified theories, the extra 
dimensions  are  usually assumed  to be spacelike. However, there 
is no {\em a priori} reason why  extra dimensions cannot be timelike. 
As a matter of fact, the consideration of extra timelike  dimensions 
in physics has a long and distinguished history \cite{Dirac}, \cite{Kastrup}, 
\cite{Salam}, \cite{Sakharov} and currently it is a subject of considerable 
interest.

In this section we critically review some objections commonly 
raised against the timelike signature of the extra coordinate. 
We argue that these objections, which are usually discussed in
 the context of compactified extra dimensions, are less clear than
 is sometimes claimed. We present arguments showing that, based on 
our current knowledge and understanding, models having a  large 
timelike extra dimension cannot  be dismissed as mathematical curiosities 
in non-physical solutions. 

\subsection{Closed timelike curves}

A common objection against extra  timelike  dimensions is 
that they  would lead to closed timelike curves (CTC), raising 
the question of time-travel and its associated paradoxes. The 
standard wisdom is that CTC appear in non-physical solutions. In 
this subsection we  discuss some recent studies which advise a serious 
reexamination of this argument and require a more receptive attitude toward CTC.

\subsubsection{Understanding CTC in $4D$ general relativity}

In four-dimensional general relativity, many solutions of 
Einstein's field equations contain CTC. The question of whether 
CTC violate the causality principle has been investigated \cite{Compatible}. 
Although the answer to this question does not seem to be  a conclusive one yet, 
CTC are usually dismissed on the grounds that the spacetimes in which they 
arise are non-physical. 

In a remarkable recent paper, Bonnor shows that 
there are a number of simple physical systems, which 
might  occur in the laboratory, or in astrophysics, where 
CTC cannot be avoided \cite{Bonnor}. One of these systems 
consists of a magnet and a static charge placed on magnet's 
axis. The solution of the Einstein-Maxwell equations for this 
system demonstrates that  CTC must occur near the axis. The 
present understanding of CTC in  general relativity does not 
give a satisfactory account of this. Bonnor confronts the usual 
time-travel interpretation of CTC by remarking that such a simple 
system {\em is not} a time machine. What this exposes is the urgent 
necessity of  finding a convincing physical  interpretation of CTC 
appearing in realistic systems. Bonnor's examples constitute 
vigorous arguments promoting that  physics can be compatible with CTC.  

 Thus, CTC can and do appear in physical solutions. The open question 
is their significance. The lesson we learn from this is that CTC can no longer 
be dismissed as {\em mathematical curiosities} occurring in non-physical solutions. 

\subsubsection{CTC in manifolds with extra dimensions}

The occurrence of CTC in GR suggests that they  also appear  in solutions in $5D$, 
even in the case of a spacelike extra dimension. The significance of CTC 
in manifolds with extra dimensions seems to be still more elusive than in $4D$. 

Indeed, in the Randall-Sundrum brane-world scenario and other 
non-compact Kaluza-Klein theories,  the motion of test particles 
is higher-dimensional in nature. In other words, all test particles 
travel on five-dimensional geodesics but observers, who are bounded to 
spacetime, have access only to the $4D$ part of the trajectory. Therefore, 
even if the $5D$ manifold contains CTC, does this imply that the 
corresponding $4D$ timelike curve in ``physical" spacetime is also a CTC?. 
In default of a reasonable interpretation of  CTC in $4D$, what can they 
mean in five-dimensional solutions? The answers to these questions appear 
to be convoluted. Specially if we take into account that a test particle 
moving geodesically in the five-dimensional manifold is perceived in $4D$ 
to be moving under the influence of an extra force, {\em not} along a 
geodesic in $4D$. 

With regard to manifolds with extra {\em timelike} dimensions, there are 
at least two more questions. Firstly, it is uncertain whether 
such manifolds inevitably give rise to CTC. Secondly, does the 
presence of two timelike dimensions necessarily cause problems with causality? 

Regarding causality,  one should be careful to discriminate between 
{\em temporal} dimensions, which actually have physical units of time;
 and {\em timelike} ones, which merely have timelike signature \cite{Wesson}.
 This argument  is supported by a recent study of the consequences of $5D$ 
relativity with two timelike dimensions \cite{Twotimes}. Also, the 
examination  of exact solutions to the field equations in $5D$ sheds 
some light on the above-mentioned  questions.

Consider for example  the class of spherically symmetric 
(in ordinary three-dimensional space) static solutions found by Billiard and Wesson 
\cite{Billiard}. These involve  $5D$ manifolds with a large extra timelike dimension. 
However, the solutions  exhibit   good physical properties. Namely (i) 
they represent centrally condensed clouds with density profiles similar
 to those of cluster of galaxies, and (ii) the analysis of the geodesic 
motion in \cite{Billiard} shows no evidence of CTC. 

\medskip

This example shows that  an extra timelike coordinate 
(at least a large one) does not rule out  the possibility of an acceptable 
physical interpretation in $4D$. Neither it automatically leads to problems 
with causality.  The absence of CTC is explained as  a consequence of that 
one of the timelike coordinates is a temporal dimension, while the 
``second" timelike coordinate is related to the inertial mass of 
the test particle, in both induced matter and brane theory 
\cite{Twotimes}. 

\medskip

The above results are quite constructive and encourage more 
future work in the areas of CTC and ``two-time" metrics. Faced with 
this situation, it is probably wise to keep an open mind toward a 
timelike signature of the extra dimension. This is the current attitude 
in theories with more than one timelike dimension, notably  in relation 
to even higher-dimensional extensions of general 
relativity \cite {Chaichian}, \cite{Yutaka}, \cite{Tianjun}, \cite{Gogber1}, 
\cite{Notachions}, \cite{Gogber2}. In string theories,   it 
is remarkable that dualities can change the number of time 
dimensions, giving rise to exotic spacetime 
signatures \cite{dualities}, \cite{signature}.  

 \subsection{Timelike extra dimensions}
In this subsection we discuss some  classical and quantum aspects of the 
dynamics of particles moving in $5D$ manifolds. We will concentrate our attention 
on manifolds with a timelike extra dimension, which is the case under scrutiny. 
We will see that there is  no evidence that a timelike large 
extra dimension {\em automatically} leads to  non-physical features.

\subsubsection{Compactified timelike extra dimension}

If, in analogy with extra spacelike coordinates,  the extra timelike dimension 
is assumed to be compactified, then the Kaluza-Klein excitations are perceived, 
by a four-dimensional observer, as states with imaginary masses, i.e., 
tachyonic states with masses quantized in units of $i |n|/\rho$, 
where  $n = 0, 1, 2, 3...$ and $\rho$ is the scale parameter 
or ``radius" of the fifth dimension. As a result of the 
tachyonic nature of the graviton KK modes, an imaginary 
part is induced in the effective low-energy (Newton's) 
gravitational potential between two test point masses \cite{Chaichian}.  
A similar situation occurs with the gravitational self-energy of massive 
bodies, in general. Also, the effective potential between two test 
charges turns out to be complex \cite{stability}. Such complex 
contributions to the energy can be associated with matter 
instability \cite{screening effect}. The disappearance of particles 
into ``nothing" would lead to problems with the conservation of charge 
and energy and contradictions with current observations \cite{massdisappearance}. 
The gravitational instability in quantum theory is another common 
objection against extra timelike dimensions.  

Because of these fundamental problems extra timelike 
dimensions {\em cannot} be hidden away by compactifying them in little 
circles, like extra spacelike dimensions in compactified KK theory. A 
possible  solution to these problems is provided by the two-time theory 
promoted by I. Bars  \cite{Bars1}. The theory possesses a new gauge 
symmetry that removes all the ghosts and overcomes the problems of  
causality and unitarity \cite{Bars1}, \cite{Bars2}. 

Thus, there seems to be no general agreement regarding  
the stability problem  in theories with  compact timelike 
extra dimensions; it   is less clear than is sometimes claimed 
\cite{massdisappearance}, \cite{Noghosts}. 

\subsubsection{Large timelike extra dimension}

Since the existence of tachyons and the consequent instability follow
 from the compactification of the extra timelike dimension, the 
applicability  of the above arguments to non-compactified timelike 
extra dimension seems to be dubious.

The question of whether tachyonic states still persist in 
quantum theory with large timelike extra dimensions is an open one. 
However, there are some preliminary results suggesting  that the 
answer to this question is negative. 

Let us first notice that from the analysis of the geodesic 
motion in $5D$, it follows that the rest mass\footnote{The rest mass in $4D$ 
is defined using the Hamilton-Jacobi formalism. The definition is independent 
of the coordinates and any parameterization used along the motion. For 
simplicity, the warp factor $\Omega$ of \cite{DynKK} is 
taken as $\Omega = 1$.} $m_{0}$ of test particles, as 
observed in $4D$,  is given by  \cite{DynKK}
\begin{equation}
\label{relation between the rest mass in 4D and 5D}
m_{0} = M_{(5)}\left[1 + \epsilon \Phi^2 \left(\frac{dy}{ds}\right)^2\right]^{- 1/2}, 
\end{equation}
where $M_{(5)}$ is the constant five-dimensional mass of the particle.
This equation shows how the motion along $y$ affects the rest 
mass measured in $4D$. It is the five-dimensional counterpart 
to  $m = m_{0}[1 - v^2]^{-1/2}$, for the variation of particle's 
mass due to its motion in spacetime. The behavior of $m_{0}$ depends 
on the signature of the extra dimension. 

For a timelike extra dimension $(\epsilon = + 1)$, the observed $4D$ rest 
mass decreases as a consequence of motion along $y$. Therefore, it cannot 
take arbitrary large values, i.e.,
\begin{equation}
\label{range of mass in 5D with timelike extra dim}
 0 < m_{0} \leq  M_{(5)}.
\end{equation}
For spacelike extra dimension $(\epsilon = -1)$, it is the opposite and 
\begin{equation}
\label{range of mass in 5D with spacelike extra dim}
M_{(5)}\leq m_{0} < \infty.
\end{equation}
If the trajectory in $5D$ is confined to hypersurfaces $y = constant$, 
then $m_{0} = M_{(5)}= const.$ along the motion. In particular, 
a massless particle in $5D$ is observed as a massless particle in $4D$. 
Also, a timelike extra dimension puts no restriction on $(dy/ds)$, 
while for a spacelike $|dy/ds| < 1 /|\Phi|$. 

Next it is possible to show \cite{QMKK} that $m_{0}$ satisfies the 
equation\footnote{These are equations (42)-(43) in \cite{QMKK}. A 
different approach allows to obtain (\ref{oscillation of mass}) 
without resorting to  the charge. } 
\begin{equation}
\label{oscillation of mass}
\frac{d^2m_{0}}{dw^2} + \epsilon m_{0} = 0,
\end{equation}
where $w = (1/2)\int{[(\partial g_{\mu\nu}/\partial y)u^{\mu}u^{\nu}]ds}$. 
This equation for $\epsilon = +1$ is the harmonic oscillator 
for $m_{0}$. In this case, by virtue of (\ref{range of mass in 5D with 
timelike extra dim}) we can write 
\begin{equation}
m_{0} = M_{5}|cos(w - \bar{w})|,
\end{equation}
where $\bar{w}$ is a constant of integration which can be expressed 
in terms of the initial values\footnote{In the case of $\epsilon = -1$, 
the solution will be $m_{0} = M_{5}cosh(w - \bar{w})$, which agrees 
with (\ref{range of mass in 5D with spacelike extra dim}).} 

\medskip

A corollary of this, regarding the question of interest here,  is that 
in manifolds with a large timelike extra dimension the observed $4D$ rest 
mass of test particles  ``oscillates" remaining always finite and 
positive\footnote{The variation of $m_{0}$ is an effect of the large 
extra dimension, $(\partial g_{\mu\nu}/\partial y) \neq 0$. It takes 
place on cosmic timescales, so it would not be observed directly in the laboratory}. 

\medskip

The above description is classical. How can we extend it to the quantum domain? 
 What is the quantum counterpart of (\ref{oscillation of mass})? In quantum 
theory the mass of a particle is not defined {\em a priori}. In a recent 
paper Wesson discusses, among other things, the question of whether the 
wave nature of particles can be understood as a manifestation 
of an $N(> 4)$-dimensional space \cite{DynWesson}. He conjectures 
that the classical and quantum dynamics in $4D$ are different 
descriptions of $5D$ dynamics in what he terms the Einstein and 
Plank gauges. This conjecture means that the $4D$ Klein-Gordon equation 
for a relativistic particle with zero spin and finite mass should be 
derivable from the $5D$ equation for a null geodesic. Without going 
into technical details, he argues that this can be done in $5D$ manifolds 
with {\em two} timelike coordinates. He obtains  an equation which is 
similar to our (\ref{oscillation of mass}) and concludes that the 
effective $4D$ mass of the particle associated with the wave 
oscillates, but the square of the mass is always finite and positive. 

\medskip

Wrapping up this part of the  discussion, we have 
\begin{enumerate}
\item At the classical level there is no contradiction between two-times 
and positivity of $4D$-effective rest mass. 
\item In the quantum description  built on  Wesson's conjecture 
there are no tachyonic states in the case of one large extra timelike 
dimension. So there is no conflict with stability.
\item The above classical  and quantum descriptions in $4D$ are totally compatible. 
\end{enumerate}

 Thus, the  classical and quantum arguments summarized above raise no 
objections against models in $5D$ manifolds with a large timelike extra 
dimension. 

What this indicates is that notions and concepts which are valid in 
compatified theory are not necessarily  valid and/or applicable in 
non-compactified theory. We would like to illustrate this  point with 
another example that comes from the so-called fifth 
force.\footnote{This is the non-gravitational force perceived by an 
observer in $4D$ who describes the motion of a test particle moving 
geodesically in $5D$.} Namely, if we apply in non-compactified theory 
the definition of force successfully used in compactified theory, then 
we get a quantity (say $f_{\mu}$) which has {\em bad} mathematical and 
physical properties. From a mathematical 
viewpoint $f_{\mu} \neq g_{\mu\nu}f^{\nu}$ and $f_{\mu}u^{\mu} \neq f^{\mu}u_{\mu}$. 
That is, this quantity is {\em not} a four-vector. From a physical viewpoint,  
it is not gauge-invariant and  the mass and/or its variation are not 
appropriately implemented.

\subsection{Four-dimensional interpretation}

  In this subsection we concentrate our attention on  the features 
that an ``acceptable"  matter distribution should 
satisfy,\footnote{It should be reiterated that the 
concepts ``reasonable" and ``acceptable" regarding the 
properties of a physical system have considerably evolved 
over the years. We already mentioned the problem with CTC. We 
should now add the question of the energy conditions, as discussed 
by Visser and Barcelo \cite{WB}} and discuss the question of how the 
signature of an extra dimension can influence the  effective gravity 
in $4D$. Our aim is to show that the physical conditions, imposed on 
the $4D$ effective matter, do not preclude the existence of a large 
timelike extra dimension. 

 \subsubsection{$4D$ effective gravity }

The $4D$ effective theory of  gravity with a  {\em compactified} 
extra dimension is usually obtained from the variation of the five-dimensional 
Einstein action. The fifth dimension is integrated out by 
virtue of the ``Kaluza-Klein ansatz", which  in practice 
allows us to drop all derivatives with respect to the extra 
coordinate $y$, as well as  to pull $\int{dy}$ out of the action 
integral. With this simplification, the original action separates 
into three pieces; these are just the actions for 
gravity\footnote{The actions for gravity and 
electromagnetism are scaled by factors of $\Phi$.}, 
electromagnetic field and massless scalar field.  

This simple procedure {\em does not} work in non-compactified theories. 
 The fifth dimension  cannot be integrated out because  of the  
explicit dependence of the five-dimensional metric on the extra coordinate. 
In this case the $4D$ effective theory of gravity is obtained directly from 
the dimensional reduction of the Einstein equations in $5D$. 

Without going into details the effective equations for gravity 
in $4D$ are \cite{NewVar}, \cite{FRLW} 
\begin{equation}
\label{4D effective theory}
{^{(4)}G}_{\alpha\beta} = \frac{1}{2}k_{(5)}^2\Lambda_{(5)}g_{\alpha\beta} 
+ 8 \pi GT_{\alpha\beta}^{(eff)},
\end{equation}
where 
\begin{equation}
\label{effective EMT in terms of extrinsic curvature}
8 \pi GT_{\alpha\beta}^{(eff)} \equiv  
- \epsilon\left(K_{\alpha\lambda}K^{\lambda}_{\beta} 
- K_{\lambda}^{\lambda}K_{\alpha\beta}\right) 
+ \frac{\epsilon}{2} g_{\alpha\beta}\left(K_{\lambda\rho}K^{\lambda\rho}
 - (K^{\lambda}_{\lambda})^2 \right) - \epsilon E_{\alpha\beta}. 
\end{equation}
This equation clearly shows that the nature of $4D$ effective 
matter depends on the signature  of the extra dimension. In order 
to get another perspective we substitute the extrinsic curvature 
from (\ref{extrinsic curvature}) into (\ref{effective EMT in terms 
of extrinsic curvature}). We obtain 
\begin{equation}
\label{Energy Momentum Tensor in STM}
 8 \pi GT_{\alpha\beta}^{(eff)} = \frac{\Phi_{\alpha;\beta}}{\Phi}
 - \frac{\epsilon}{2\Phi^2}
\left[\frac{\stackrel{\ast}{\Phi} \stackrel{\ast}{g}_{\alpha \beta}}{\Phi} 
- \stackrel{\ast \ast}{g}_{\alpha \beta} 
+ g^{\lambda\mu}\stackrel{\ast}{g}_{\alpha\lambda}\stackrel{\ast}{g}_{\beta\mu} 
- \frac{1}{2}g^{\mu\nu}\stackrel{\ast}{g}_{\mu\nu}\stackrel{\ast}{g}_{\alpha\beta}
 + \frac{1}{4}g_{\alpha\beta}\left(\stackrel{\ast}{g}^{\mu\nu}\stackrel{\ast}{g}_{\mu\nu} 
+ (g^{\mu\nu}\stackrel{\ast}{g}_{\mu\nu})^2\right)\right],
\end{equation}
where $\stackrel{\ast}{f} = \partial f/\partial y$.  
It shows that the signature of a compact extra dimension 
(for which $\stackrel{\ast}{f} = 0$) does not affect the nature of 
the $4D$   effective matter;  it is radiation-like (because $E_{\mu\nu}$ is traceless) 
for any model in $5D$. However, if the extra dimension is 
large ($\stackrel{\ast}{f} \neq 0$), then  the signature  
crucially affects   the interpretation of $4D$ matter .

\subsubsection{Conditions on the effective matter}

In STM the effective stress-energy tensor $T_{\alpha\beta}^{(eff)}$ 
is commonly assumed to be a perfect fluid. Then physical ``restrictions", 
as the energy conditions and  an equation of state,  are imposed on the 
effective density and pressure. 

In brane theory, with the introduction of ${\bf Z}_2$ symmetry 
about our brane-universe, $T_{\alpha\beta}^{(eff)}$  is interpreted 
as the sum  of the energy-momentum tensor on the brane $\tau_{\mu\nu}$ 
plus local and non-local (Weyl) corrections. Using (\ref{K in terms of S})
 and (\ref{decomposition of tau}), the $4D$ effective field equations  
(\ref{4D effective theory}) become
\begin{equation}
\label{EMT in brane theory}
^{(4)}G_{\mu\nu} =  {\Lambda}_{(4)}g_{\mu\nu} + 8\pi G T_{\mu\nu} 
- \epsilon k_{(5)}^4 \Pi_{\mu\nu} - \epsilon E_{\mu\nu},
\end{equation}
where ${\Lambda}_{(4)}$, $G$ and $E_{\mu\nu}$ are 
given by (\ref{definition of lambda}), (\ref{effective gravitational coupling})
 and (\ref{Weyl Tensor}), respectively. The symmetric tensor 
$\Pi_{\mu\nu}$ represents the quadratic local corrections, viz., 
\begin{equation}
\label{quadratic corrections}
\Pi_{\mu\nu} =  \frac{1}{4} T_{\mu\alpha}T^{\alpha}_{\nu} 
- \frac{1}{12}T T_{\mu\nu} - \frac{1}{8}g_{\mu\nu}T_{\alpha\beta}T^{\alpha\beta} 
+ \frac{1}{24}g_{\mu\nu}T^2.
\end{equation}
All these four-dimensional quantities have to be evaluated on $\Sigma^{+}$. 

In addition to  the energy conditions \cite{WB} on the energy-momentum
 tensor on the brane $T_{\mu\nu}$, the physical interpretation in $4D$
 requires the positiveness of $G =  ( - \epsilon \sigma){k_{(5)}^4}/{48 \pi} $. 

Thus, in the brane-world scenario a spacelike extra dimension requires 
the vacuum energy $\sigma$ to be  positive, while a timelike extra dimension 
requires $\sigma$ to  be negative. Notice that if $\Lambda_{(4)} = 0$, 
then $\Lambda_{(5)} = \epsilon k_{(5)}^2 \sigma^2/6$. Thus, a vanishing
 cosmological constant in $4D$ establishes  a link between  the signature
 of the extra dimension and the sign of the cosmological constant in the bulk.
 For a spacelike (timelike) extra dimension, the bulk must be $AdS_{5}(dS_{5})$.

\subsubsection{Examples of $5D$ solutions and their interpretation  in $4D$}

In principle, any solution of  Einstein's field equations 
in $5D$ can be interpreted either in the context of STM or 
brane theory \cite{STMBrane}. We assume that a phenomenologically acceptable 
 $4D$ effective theory should satisfy the physical conditions mentioned 
above.\footnote{Probably not every solution of the five-dimensional 
equations generates an effective matter in  $4D$ spacetime with  
physically reasonable density and pressure, as well as $G > 0$. Just as 
in $4D$ general relativity where not every solution 
of the field equations is a physical solution representing 
 a situation which might occur in astrophysics, or in cosmology.
 Here we  dismiss those solutions as non-physical.} This
 assumption does not compromise the nature of the extra 
dimension.  Indeed, there are several  models which show 
reasonable matter distributions in $4D$, and involve $5D$ 
manifolds with  $\epsilon = -1$, $\epsilon = +1$, or both. 
They fall into one of the following categories.

(1) Models which exist  only for one signature. As an example 
we mention the five-dimensional cosmological model with metric coefficients 
 that are separable functions of $t$ and $y$. The field equations in $5D$ 
lead to a class of  solutions that only exists for $\epsilon = -1$. These
 solutions embed the flat FRW cosmologies and exhibit  good physical 
properties \cite{JPdeL 1}. Another example of this kind is provided by 
the ``wave-like"  model discussed in \cite{Billiard2}. That model is 
distinct from the one discussed here.  The corresponding $5D$ field 
equations have solution only for $\epsilon = + 1$, and the metric 
coefficients for the ordinary $3D$ space are  complex.  This is 
quite out of the ordinary in relativity.   However, the $4D$ effective
 physical quantities are real (We do not want to discuss the meaning of 
complex metrics here, for this see \cite{Billiard2}. We mention it as a 
concrete example of solutions that exist only for $\epsilon = +1$).

(2) Models which exist for both signatures, but work  
properly only with one of them. A nice example of this is given 
by the static spherical (in $3D$ space) model with metric coefficients 
 that are separable functions of $r$ and $y$ \cite{Billiard}. The 
field equations in $5D$ yield the static solution that  we have 
discussed in Sec. $5.1.2$. In principle both signatures $(\epsilon = \pm 1)$ 
are possible. However, the matter in $4D$ spacetime has physically reasonable 
density and pressure only  for $\epsilon = + 1$. Another example of this kind 
is provided by the model where  $g_{00} = 1$. It is  discussed in \cite{NewVar}. 
Once again the field equations have solutions for both signatures. However, 
only the solutions with $\epsilon = -1$ present good physical properties.
 Any attempt of extending the validity of these solutions to an  extra 
dimension with  the opposite signature, leads to contradictions like 
negative mass in the first example and negative $G$ in the second one. 

(3) Models which exist, and work  properly,  for  both signatures.
The wave-like model discussed here exemplifies this case. 
The solution in section  $3.2.1$ has $\epsilon = \pm 1$ so 
the extra dimension can be spacelike or timelike. Besides, 
 the $4D$ effective  matter has physically reasonable density 
and pressure, along with positive $G$, for both signatures. 
However, this is {\em not} just a mathematical extension of the 
validity of the solution from one signature to another. The physical 
properties of solutions with a timelike extra dimension are very 
different from the ones with spacelike extra dimension. The signature 
affects not only the matter distribution, but also the motion of test 
particles. In fact, the cosmological constant in the bulk, the vacuum 
energy, the dynamical evolution of the universe, the masses,  and the 
motion of test particles,  in manifolds with  $\epsilon = + 1$ are
 radically distinct from those in manifolds with  $\epsilon = -1$. 
Another example with similar properties is the model having $\Phi = 1$, 
discussed in \cite{NewVar}. It includes the Randall-Sundrum scenario with 
an extra timelike  coordinate, instead of a spacelike one \cite{Chaichian}.

\medskip

The analysis of this section leads us to conclude  
that the physical conditions, imposed on the $4D$ effective matter, 
do not preclude the existence of a large timelike extra dimension. 

\section{Summary and conclusions}

We have studied brane-world cosmologies embedded in a bulk where 
the five-dimensional  metric functions are plane-waves propagating 
along the extra dimension. The motivation for this has been to model 
the singular character of the brane as a result of the collision of 
waves moving in opposite directions along $y$. At the plane of collision 
the metric is continuous and  the derivatives with respect to $y$, 
calculated at each side of the plane, are equal in magnitude but have 
opposite sign. Thus, the ${\bf Z}_2$ symmetry used in brane-world theory 
is inherent to the model.

As a consequence  of the  wave-like nature of the metric, the 
Einstein equations in $5D$ reduce to a set of two {\em ordinary} 
differential equations, (\ref{Phi}) and (\ref{relation between a and n}), 
for determining the metric functions  $n$, $a$ and $\Phi$. Therefore we have 
to complete the system of equations by making some suitable assumption. In 
Section 2.2 we showed a quite general solution that indicates that the $5D$ 
equations  admit interesting wave-like solutions without imposing severe 
restrictions on the model. 

In Section 3 we used the brane-world paradigm to formulate the 
appropriate physical assumptions to complete the model. We assumed 
the isothermal equation of state for ordinary matter on the brane, 
and obtained an equation that links $n$ and  $a$ with the tension $\sigma$ 
of the brane (\ref{relation between n, a and sigma}). So we still need 
another assumption. The simplest one is to consider that the tension is 
constant, although there are physical models where this is not a viable 
assumption \cite{NewVar}. This completes the specification of the model.

We have shown that the model gives back the generalized Friedmann
equation (\ref{generalized FLRW equation}) for the cosmological 
evolution on the brane, although it is not possible to solve the 
five-dimensional equations exactly for an arbitrary set of the 
parameters appearing in the theory. Except for the explicit 
dependence on $\epsilon$, this is the same equation obtained 
previously from models with static $(\dot{\Phi} = 0)$, spacelike fifth 
dimension. It is important that we recover the familiar evolution equations 
in $4D$ for a wide variety of cosmologies and settings in $5D$. 

The signature of the extra dimension comes out in the quadratic correction, 
in front of $\rho^2$. Consequently, its effects are important at early 
stages of the evolution,  when this correction becomes dominating. 
Setting $\Lambda_{(4)} =0$, we find that models with a timelike 
extra dimension show a bounce at some finite $a_{min}$, where 
the geometry is regular and the energy density is nonsingular. This 
is opposed to the big bang solutions, for a spacelike extra dimension, 
where the geometry suffers a breakdown and the energy density diverges. 
In both  cases, we recover the physics of the late universe 
for $(\sigma >> \rho)$. 

Models with bounce can happen in general relativistic FLRW models 
that have a large, positive, cosmological constant. In contrast here 
we have $\Lambda_{(4)} = 0$, and the bounce is a genuine product of 
the timelike extra dimension.  

The Weyl tensor in the bulk affects the value of $a_{min}$ through $\beta$. 
We have shown that for arbitrary large and positive values of $\beta$, the 
bouncing ``radius"  $a_{min}$ can be as near as one wants to $a = 0$. In any 
case, the Weyl tensor does not affect the overall character of the solutions.

We also discussed, in Section $3.2.2$, the five-dimensional 
wave-like solution with static fifth dimension,  $\dot{\Phi} = 0$. In $4D$ 
it corresponds to Milne's universe. The brane is not empty and the tension 
turns out to be a function of $a$. Consequently, the resulting $G$ 
and $\Lambda_{(4)}$  are not constants but vary with the evolution 
of the universe. As we mentioned earlier, we have studied other 
models with similar behavior elsewhere \cite{NewVar}.

Our model also predicts the development  of the extra dimension. In 
Section 4, we have seen that the dynamics  of $\Phi$ is influenced by 
the matter on the brane through its influence on the expansion rate. 
The main features of $\Phi$ at early stages of the evolution  are 
significantly affected by the signature of the extra dimension, 
although the late behavior is the same in both cases.  On the basis 
of our model we can reach some general conclusions.

(i) Although $\Phi$ is small today, it is {\em growing} in size if 
the universe is speeding up its expansion. The opposite also holds, 
the size of $\Phi$ is decreasing is the universe is speeding down its 
expansion. 

(ii) The relative change of $\Phi$ is determined by the Hubble
 and deceleration parameters as shown in (\ref{rate of change of Phi}). 

(iii) At any time during  the 
evolution\footnote{The parameter $\alpha$ can be taken
 as $\alpha = 1$, without loss of generality.} $(\alpha \Phi) = H a$. 

We reiterate that these conclusions are general in the sense 
that they are the same regardless  of the details of the model 
in $4D$, i.e., the value of $k, C, \Lambda_{(4)}, \gamma, \sigma$ and $\beta$.  
We note that the evolution  of the extra dimension is 
 independent of whether we choose  the brane world paradigm 
or STM to recover the $4D$ effective gravity. However, the 
dynamics of our four-dimensional universe does depend on this
 choice \cite{FRLW}.

 In order to avoid  misunderstanding we would like to 
emphasize that the above discussion refers to the case 
where the bulk is filled with only a cosmological 
constant, i.e.,  ${^{(5)}T}_{AB} =  \Lambda_{(5)}g_{AB}$. In 
this case, the relation $(d\Phi/d\tau_{\Sigma})/\Phi = - q H$ is a 
direct consequence of $\dot{a} = \alpha n \Phi$, which follows  
from $G^{0}_{4} = 0$.

In general, from the field equation $G_{4\mu} = k_{(5)}^2 {^{(5)}T}_{4 \mu}$ 
and (\ref{emt on the brane in terms of K}) it follows that 
\begin{equation}
\tau^{\mu}_{\nu;\mu} = - \frac{2 \epsilon}{\Phi}{^{(5)}T}_{4 \nu},
\end{equation}
and 
\begin{equation}
G_{4 \nu} = - \frac{\epsilon k_{(5)}^2 \Phi}{2}\tau^{\mu}_{\nu;\mu}.
\end{equation}
Consequently,  in the case where the bulk contains scalar and/or other 
fields, so that ${^{(5)}T}_{4\nu} \neq 0$, the brane energy-momentum
 tensor $\tau_{\mu\nu}$ is not conserved. As a result of this $G_{4\nu} \neq 0$ 
and therefore $(d\Phi/d\tau_{\Sigma})/\Phi = - q H$ does not hold. In 
other words, the conclusions (i)-(iii) above are valid only if the
 brane energy-momentum tensor is conserved.  

We would like to finish this paper with two more comments:

The first one is related to the question of whether the solutions 
discussed here, with $\epsilon = - 1$,  are isometric to the 5-dimensional 
Schwarzschild-$AdS_{5}$
topological black hole.   
In $4D$ general relativity, it is well known that the Schwarzschild metric is the unique spherically
symmetric solution of vacuum Einstein's equations. Unfortunately, in $5D$ this result is not 
valid in general. Birkhoff's theorem holds  only for the case of static $\Phi$ $(\dot{\Phi} = 0)$ in  metric (\ref{cosmological metric}). Indeed, in   this case there 
exist a coordinate transformation \cite{mukoyama2} which links the $5D$ line element of the 
Schwarzschild-de Sitter bulk ($\Phi \neq 1$, but $\dot{\Phi} = 0$) with the ``geodesic" bulk $(\Phi = 1$ and $\dot{\Phi} = 0)$. In the case of non-static $\Phi$ $(\dot{\Phi} \neq 0)$, there are a number of non-stationary solutions to 
the vacuum Einstein equations in $5D$ \cite{WLPdel}-\cite{kokarev},  which are {\em not} equivalent to the 5-dimensional Schwarzshild-AdS spacetime. Such solutions generally provide new insight into some important problems. In particular, the five-dimensional model discussed here can be used as an embedding for cosmologies with variable physical ``constants" \cite{GLH}. 

The second comment is related to equation (\ref{turning points}), which   
implies that there is some sort of
singularity in the metric whenever $da/d\tau_\Sigma = 0$, since $\Phi$
disappears there. The good news is that this singularity is not directly measurable by an observer in $4D$, for whom  all physical quantities are finite at all times $(\epsilon = + 1)$.  The bad news is that  we  can do nothing about it, irrespective of whether this is a coordinate or curvature singularity. Certainly, if this is a coordinate singularity  it can be suppressed by a transformation of coordinates in $5D$. From a mathematical point of view the ``old" metric and the transformed one represent the same five-dimensional manifold. However, from a four-dimensional viewpoint they are not equivalent: they give rise to different scenarios in $4D$. The structure and the material content of the observed spacetime is changed by any transformation that involves the extra coordinate. This is similar to four-dimensional physics, where a coordinate transformation involving time also implies a change in the system of reference, and consequently also modifies the observed physical picture.  This is an important point and we should come back to it in a future publication.

\end{document}